\documentclass[twocolumn,prl,floatfix,citeautoscript,nofootinbib,superscriptaddress]{revtex4-2}
\usepackage{amsbsy}
\usepackage{latexsym,epsfig,graphicx}
\usepackage{dcolumn}
\usepackage{graphicx}
\usepackage{subfigure}
\usepackage{comment}
\usepackage{color}
\usepackage{bm}
\usepackage{mathrsfs}
\usepackage{amsfonts}
\usepackage{amsmath}
\usepackage{color}
\usepackage{amssymb}
\usepackage{xspace}
\usepackage{epstopdf}
\usepackage{tabularx}
\usepackage{braket}
\usepackage{longtable}
\usepackage[colorlinks=true, letterpaper=true, pdfstartview=FitV, linkcolor=blue, citecolor=blue, urlcolor=blue]{hyperref}
\usepackage[normalem]{ulem}

\pdfoutput=1

\begin{document}

\title{Universal intrinsic higher-rank spin Hall effect}
\author{Junpeng Hou}
\altaffiliation{These authors contributed equally to this work}
\affiliation{Department of Physics, The University of Texas at Dallas, Richardson, Texas
75080-3021, USA}
\author{Ying Su}
\altaffiliation{These authors contributed equally to this work}
\affiliation{Department of Physics, The University of Texas at Dallas, Richardson, Texas
75080-3021, USA}
\author{Chuanwei Zhang}
\thanks{Email: chuanwei.zhang@utdallas.edu}
\affiliation{Department of Physics, The University of Texas at Dallas, Richardson, Texas
75080-3021, USA}

\begin{abstract}
Spin Hall effect (SHE), a fundamental transport phenomenon with non-zero
spin current but vanishing charge current, has important applications in
spintronics for the electrical control of spins. Owing to the half-spin
nature of electrons, the rank of spin current (determined by the rank of
spin tensors) has been restricted to 0 and 1 for charge and spin Hall
effects. Motivated by recent studies of pseudospin-1 fermions in solid state
and cold atomic systems, here we introduce and characterize higher-rank ($%
\geq 2$) SHEs in large spin ($\geq 1$) systems. We find a universal rank-2
spin Hall conductivity $e/{8}\pi $, with zero rank-0 and 1 conductivities,
for a spin-1 model with intrinsic spin-orbit coupling. Similar rank-2 SHEs
can also be found in a spin-3/2 system. An experimental scheme is proposed
to realize and measure rank-2 SHEs with pseudospin-1 ultracold fermionic
atoms. Our results reveal novel spin transport phenomena in large spin
systems and may find important applications in designing innovative
spintronic devices.
\end{abstract}

\maketitle

\begin{figure}[t]
\centering
\includegraphics[width=0.48\textwidth]{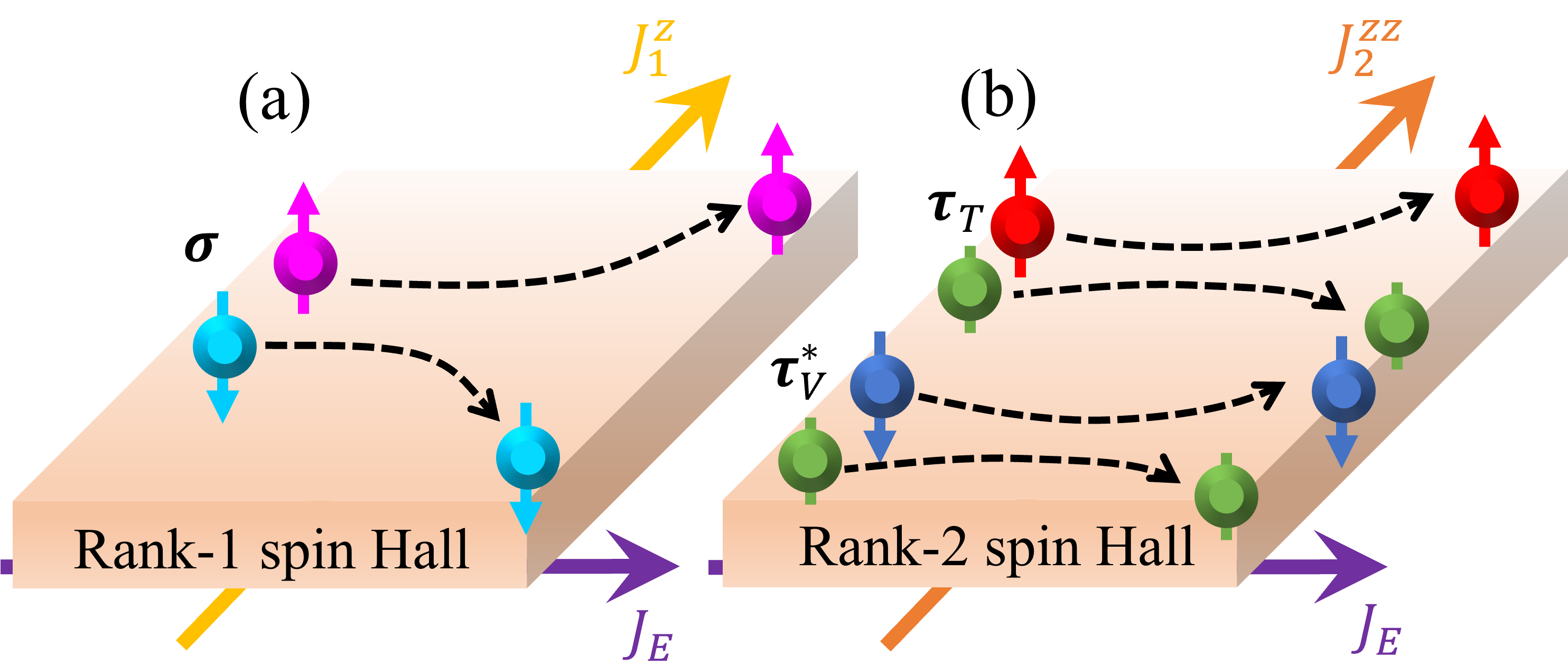}
\caption{Illustration of higher-rank spin-Hall effect. (a) In the rank-1 SHE
of a spin-1/2 system, the spin up and down components move along opposite
directions, yielding a vanishing rank-0 charge Hall current but a finite
rank-1 spin Hall current $\mathbf{J}_{1}^{z}$. (b) Rank-2 SHE in a spin-1
fermionic system, where both charge (rank-0) and rank-1 spin currents
vanish, leaving only non-zero rank-2 spin current $\mathbf{J}_{2}^{zz}$. The
red, green and blue disks correspond to spin components $|m_{z}=1\rangle $, $%
|0\rangle $ and $|-1\rangle $ respectively. In both cases, the electric
field is applied along the $x$ direction, leading to a charge current $%
\mathbf{J}_{E} $ while transverse charge current $\mathbf{J}_0$ remains
zero. }
\label{fig1}
\end{figure}

{\color{blue}\emph{Introduction}}. Hall effects and their quantized siblings
are one of the major cornerstones of modern condensed-matter physics, and
the discovery of novel Hall effects often opens new avenues for controlling
electronic transport for device applications. One of the most notable
examples in this context is probably the spin Hall effect (SHE), where spin
up and down of electrons move along opposite transverse directions under an
applied electric field, yielding non-zero spin current but vanishing charge
current (see Fig.~\ref{fig1}(a) for an illustration) \cite%
{HirschJE1999,SinovaJ2015}. Spin-current-based phenomena such as giant SHE 
\cite{SekiT2008,LiuL2012,NiimiY2012}, inverse SHE \cite%
{SaitohE2006,KimuraT2007,MiaoBF2013,RojasSanchezJC2014,Kimata2019} and
quantum SHE \cite{KaneCL2005,BernevigBA2006,Konig2007}, have also been
widely studied. SHE provides a powerful tool for controlling spins
electrically, thus has significant applications for realizing low-power
spintronic devices \cite{SinovaJ2015}.

The origin of SHE can be attributed to either extrinsic impurity scattering 
\cite{Kato2004} or intrinsic spin-orbit coupling (SOC) \cite%
{Murakami2003,SinovaJ2004}. In intrinsic SHE, the SOC serves as an effective
magnetic field that is opposite for spin up and down, yielding nonzero
spin-Hall current. For instance, in a two-dimensional (2D) electronic gas,
the Rashba SOC yields a spin-Hall conductivity $e/8\pi $, which is a
universal constant that does not depend on the underlying material
properties \cite{SinovaJ2004}. Here the spin current operator is generally
defined by $\mathbf{J_s}=\frac{1}{2}\left\{ {S}_{z},\mathbf{v}\right\} _{{+}%
} $ with $S_{z}={\frac{\hbar }{2}}\sigma _{z}$ and\ the rank-1 Pauli matrix $%
\sigma _{z}$, where $\left\{ \cdot ,\cdot \right\} _{{+}}$ denotes the
anticommutator and $\mathbf{v}$ is the velocity operator. In this sense, the
charge-current operator $\sigma _{0}\mathbf{v}$ can be viewed as rank-0, {%
where $\sigma _{0}$ is the $2\times 2$ identity matrix}. For electrons with
half spin, 0 and 1 are only available ranks for spin-1/2 matrices.

Recent theoretical and experimental advances in the study of pseudospin-1
fermions have opened a new perspective towards the realization of novel
quantum phases and dynamics in large-spin systems, in which higher-rank
spin-tensors exist and play a crucial role \cite{KawaguchiY2012}. In
particular, triply-degenerate fermions were proposed as novel quasiparticles
without counterparts in quantum field theory \cite{Bradlyn2016,HuH2018} and
certain experimental signatures have been observed in solid-state materials 
\cite{LvBQ2017}. Moreover, large-spin ($>1/2$) is easily accessible in
experiments for ultracold atoms, superconducting qubits, and trapped ions
with multiple pseudospin states \cite%
{CampbellDL2016,LuoX2016,OllikainenT2019,Tan18}, which can host interesting
quantum phases \cite%
{SunK2016,YuZQ2016,MartoneGI2016,KonigEJ2018,LuoXW2017,Murch} and
topological states \cite%
{KuzmenkoI2018,HouJ2018,PalumboG2018,PalumboG2019,Tan21}.

While these works reveal many fascinating phenomena, spin transport in large
spin systems, particularly when involve higher-rank spin tensors, remains
largely unexplored. A natural question is whether there is intrinsic
higher-rank ($\geqslant 2$) SHE where the lower-rank spin and charge
currents are zero. Here the rank of the spin current is defined through the
spin tensor matrix rank in the spin current operator. If so, can the
higher-rank spin-Hall conductivity be a universal constant independent of
material properties? Can we realize and observe higher-rank SHE in a
realistic physical system?

In this Letter, we address these important questions by defining and
characterizing \textit{universal intrinsic higher-rank SHE}, and exploring
the corresponding experimental realization using cold atoms. Our main
results are:

\textit{i)} We define spin currents of different ranks in a spin-1 system
and introduce the concept of rank-2 SHE. Higher-rank SHE for arbitrary spin-$%
F$ can be defined in a similar manner.

\textit{ii)} We develop a minimum spin-1 model for realizing rank-2 SHE with
intrinsic 2D spin-tensor-momentum coupling{\ (STMC)}. We find that the
rank-2 spin-Hall conductivity is a universal constant $e/8\pi $, which is
independent of material properties, and both rank-0 charge-Hall and rank-1
spin-Hall conductivities are zero. This is confirmed by solving both spin
dynamics and linear response theory. We also showcase another rank-2 SHE in
a spin-3/2 model.

\textit{iii)} We propose an experimental scheme for realizing the 2D STMC in
the minimal spin-1 model of rank-2 SHE. This is done by utilizing ultracold
fermions in a 2D optical lattice with STMC, which is built upon recent
experimental advances on realizations of 2D SOC for pseudospin-1/2 atoms 
\cite{WuZ2016}. We propose that rank-2 SHE can be observed by measuring
rank-2 spin accumulations \cite{Kato2004}.

{\color{blue}\emph{Rank-2 spin currents and spin-Hall effect}}. In a
spin-1/2 system characterized by the SU(2) group, the spin operators are
defined by Pauli matrices $\sigma _{i}$, which are rank-1 spin vectors
satisfying $\{\sigma _{i},\sigma _{j}\}_{{+}}=$ $\delta _{ij}$, and do not
allow rank-2 spin tensors. In a spin-1 system with the SU(3) group, the
rank-1 spin vectors $F_{i}$ do not satisfy $\{F_{i},F_{j}\}_{{+}}\propto $ $%
\delta _{ij}$, with rank-2 spin tensors defined as $N_{ij}=\{F_{i},F_{j}\}_{{%
+}}/2-\delta _{ij}\bm{F}^{2}/3$ \cite{KawaguchiY2012}. Along the
quantization axis $z$, rank-1 and 2 spin polarization operators can be
defined as $P_{1}={\hbar }F_{z}$ and $P_{2}={\hbar }N_{zz}$, leading to spin
current density operators $\mathbf{J}_{1}^{z}=\frac{1}{2}\left\{ P_{1},%
\mathbf{v}\right\} _{{+}}$ and $\mathbf{J}_{2}^{zz}=\frac{1}{2}\left\{ P_{2},%
\mathbf{v}\right\} _{{+}}$. The definition naturally yields usual rank-1
spin polarization $\left\langle F_{z}\right\rangle =\psi ^{\dagger }{\hbar }%
F_{z}\psi $ and spin current density $\left\langle \mathbf{J}%
_{1}^{z}\right\rangle =\frac{1}{2}$Re$[\psi ^{\dagger }\left\{ {\hbar }F_{z},%
\mathbf{v}\right\} _{{+}}\psi ]$, where $\psi $ is the spinor state of the
particle. Note that the charge current $\left\langle \mathbf{J}%
_{0}\right\rangle =$ Re$[\psi ^{\dagger }\mathbf{v}\psi ]$ can be treated as
the current of rank-0 unit matrix $I$.

For widely studied intrinsic universal rank-1 SHE for spin-1/2 electrons
(illustrated in Fig. \ref{fig1}(a)), the applied electric field induces
non-zero transverse currents, which are opposite for spin up and down, due
to the intrinsic SOC that serves as opposite effective magnetic fields.
Therefore the rank-0 total charge current is zero, but the rank-1 spin
current is non-zero. Similarly, we can define rank-$2$ SHE as that with only
non-zero rank-$2$ spin current (rank-1 spin and rank-0 charge currents both
vanish). The corresponding spin current configuration is illustrated in Fig.~%
\ref{fig1}(b), where both spin $\ket{+1}$ and $\ket{-1}$ move in the same
direction for vanishing rank-1 spin current, while a doubled spin $|0\rangle 
$ current flows in the opposite direction for the sake of zero charge
current. The resulting rank-2 spin current, defined through $%
N_{zz}=F_{z}^{2}-\frac{2}{3}=$diag$(1/3,-2/3,1/3)$, is clearly non-zero due
to the current directions of three spin components.

{\color{blue}\emph{Universal intrinsic rank-2 spin-Hall effect}}. A general
form of SOC in a spin-1 system may lead to non-zero spin currents with
different ranks. In order to find suitable SOC for rank-2 SHE, we adopt the
Gell-Mann matrix representation $\lambda _{i}$, $1\leq i\leq 8$ for the
SU(3) group, which can be grouped into three different SU(2) subalgebras 
\cite{Robert1984} 
\begin{equation}
\bm{\tau}_{T}=\{\lambda _{1},\lambda _{2},\lambda _{3}\},\bm{\tau}%
_{U}=\{\lambda _{4},\lambda _{5},\lambda _{+}\},\bm{\tau}_{V}=\{\lambda
_{6},\lambda _{7},\lambda _{-}\}.  \label{eq1}
\end{equation}%
in the {two-}spin subspaces $\left\{ \left\vert +1\right\rangle ,\left\vert
0\right\rangle \right\} _{T}$, $\left\{ \left\vert +1\right\rangle
,\left\vert -1\right\rangle \right\} _{U}$ and $\left\{ \left\vert
0\right\rangle ,\left\vert -1\right\rangle \right\} _{V}$. Here $\lambda
_{\pm }=\frac{\sqrt{3}}{2}\lambda _{8}\pm \frac{1}{2}\lambda _{3}$. {Spin-1
vectors are related as $F_{x}=(\lambda _{1}+\lambda _{6})/\sqrt{2}$, $%
F_{y}=(\lambda _{2}+\lambda _{7})/\sqrt{2}$, and $F_{z}=\lambda _{+}$.}
Physically, each subalgebra spans the symmetry group of an{\ effective}
quantum 
{\ spin-1/2 in the two-spin subspace and works as the Pauli matrices.}

For each SU(2) subalgebra, we consider the intrinsic SHE with Rashba SOC 
\cite{SinovaJ2004} 
\begin{equation}
H_{{\alpha }}^{\text{Rashba}}=\frac{p^{2}}{2m^{\ast }}-\frac{\lambda }{\hbar 
}\bm{\tau}_{{\alpha }}\cdot (\bm{\bm{\hat{z}}}\times \bm{p}),
\end{equation}%
where $m^{\ast }$ is the effective mass of electron, $\lambda >0$ is the
Rashba coupling strength, and 
{\ $\alpha =T,U,V$}. When the electric field $E_{x}$ is applied along the $x$
direction (Fig.~\ref{fig1}), there is a rank-1 spin Hall conductivity $%
\sigma _{xy}^{z}=\frac{e}{8\pi }$ {for $\alpha =T,V$} \cite{SinovaJ2004}. {%
Note that $\sigma _{xy}^{z}=\frac{e}{4\pi }$ for $\alpha =U$ because the
subalgebra is in the subspace spanned by $\left\{ \left\vert +1\right\rangle
,\left\vert -1\right\rangle \right\} _{U}$ with doubled spin difference 2}$%
\hbar $.

The {rank-}$2${\ spin-}Hall conductivity can be computed using the Kubo
formula 
\begin{equation}
\sigma _{xy}^{zz}=-{e\hbar }\int \frac{d^{2}\mathbf{k}}{\left( 2\pi \right)
^{2}}\Omega _{xy}^{zz}
\end{equation}%
where $\Omega _{xy}^{zz}=-\sum_{m\neq m^{\prime }}(f_{m^{\prime }\bm{k}}-f_{m%
\bm{k}})\frac{\text{Im}\langle m^{\prime }\bm{k}|J_{2,x}^{zz}|m\bm{k}\rangle
\langle m\bm{k}|\text{v}_{y}|m^{\prime }\bm{k}\rangle }{(E_{m\bm{k}%
}-E_{m^{\prime }\bm{k}})^{2}}$ is the generalized rank-2 Berry curvature, 
$m$ and $m^{\prime }$ are band indices, $f_{m\bm{k}}={\left[ e^{(E_{m\bm{k}%
}-E_{F})/k_{B}T}+1\right] ^{-1}}$ is the Fermi-Dirac distribution, the
velocity operator $\mathbf{v}=\partial _{\bm{p}}H$, {and $J_{2,x}^{zz}$ is
the $x$ component of the rank-}$2${\ spin-Hall current operator.} {Here} $%
E_{m\bm{k}}$ and $|m\bm{k}\rangle $ are eigenvalues and eigenvectors in the
momentum space. 

\begin{figure}[t]
\centering
\includegraphics[width=0.48\textwidth]{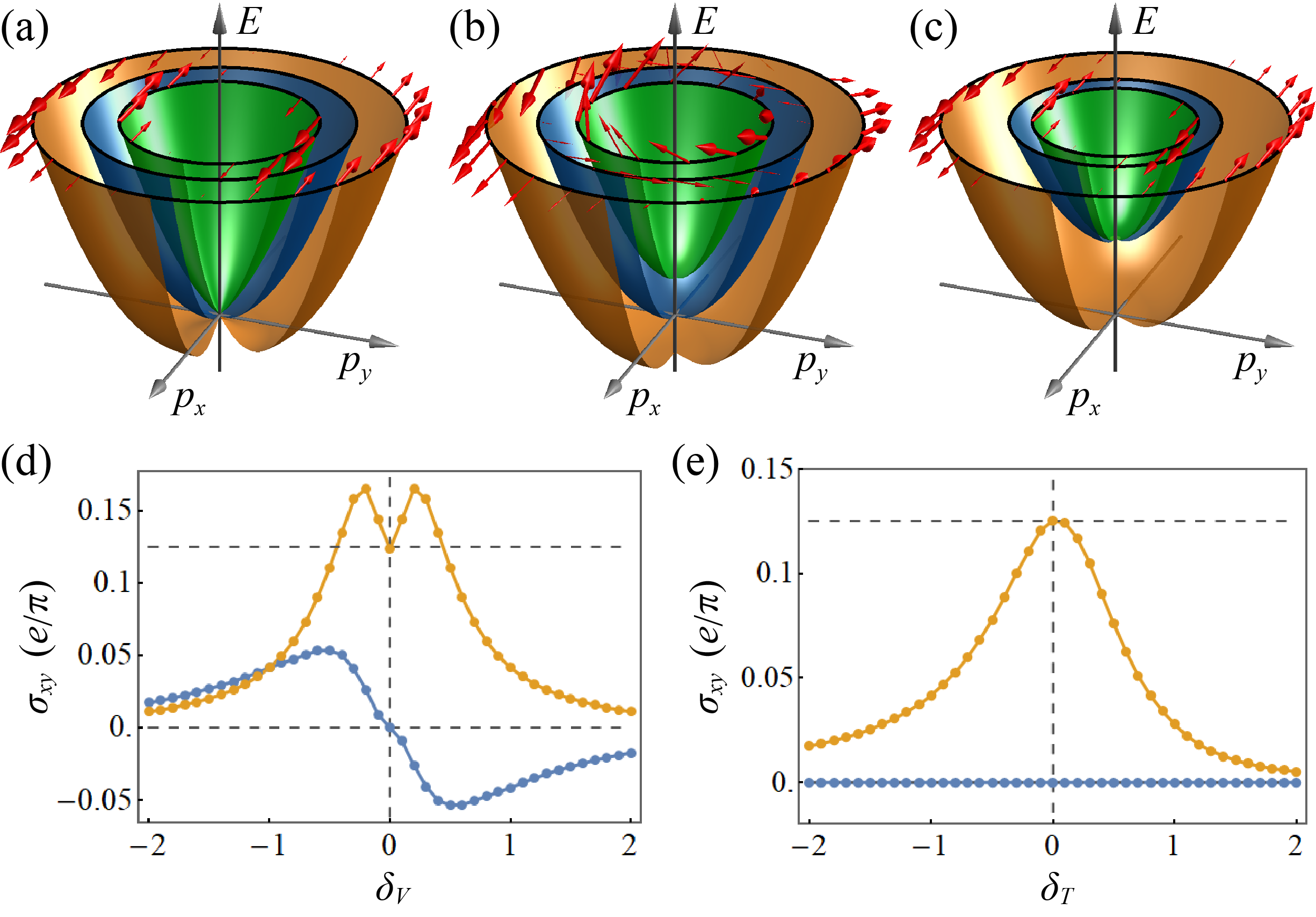}
\caption{(a-c) Representative energy bands of the STMC Fermi gas described
by Eq.~(\protect\ref{rank2H}) with (a) no Zeeman field, (b) spin vector
Zeeman field $\protect\delta _{V}F_{z}$, and (c) spin tensor Zeeman field $%
\protect\delta _{T}F_{z}^{2}$. Red arrows depict the spin textures on the
Fermi surfaces. (d) and (e) Rank-1 (blue) and rank-2 (orange) spin-Hall
conductivities with respect to vector and tensor Zeeman fields $\protect%
\delta _{V}$ and $\protect\delta _{T}$. $E_{F}=1$ and $p_{F}=1$ are taken as
units for energy and momentum. The corresponding dimensionless parameters
are set as $m=0.2$ and $\protect\lambda =0.5$.}
\label{fig2}
\end{figure}

The three SU(2) subalgebras are not independent. Through blending two of the
subalgebras, we construct the following Hamiltonian 
\begin{equation}
H_{F=1}=\frac{p^{2}}{2m^{\ast }}-\frac{1}{\sqrt{2}}\frac{\lambda }{\hbar }(%
\bm{\tau}_{T}+\bm{\tau}_{V}^{\ast })\cdot (\bm{\hat{z}}\times \bm{p}),
\label{rank2H}
\end{equation}%
where $(\bm{\tau}_{T}+\bm{\tau}_{V}^{\ast })\cdot (\bm{\hat{z}}\times \bm{p}%
)={p\left( 
\begin{array}{ccc}
0 & -ie^{-i\theta _{\mathbf{p}}} & 0 \\ 
ie^{i\theta _{\mathbf{p}}} & 0 & ie^{i\theta _{\mathbf{p}}} \\ 
0 & -ie^{-i\theta _{\mathbf{p}}} & 0%
\end{array}%
\right) }$ describes a 2D STMC. The resulting band structure is plotted in
Fig.~\ref{fig2}(a), which exhibits a 2D triply-degenerate point at $\bm{p}=0$%
. When the Fermi level lays above the triply-degenerate point, the Fermi
surfaces are simply concentric circles due to isotropic SOC, as shown in
Fig.~\ref{fig2}(a).

Applying the Kubo formula with $\mathbf{J}_{2}^{zz}=\frac{1}{2}\{{\hbar }%
N_{zz},\mathbf{v}\}$, we find a non-vanishing rank-2 spin-Hall conductivity {%
\cite{SM} } 
\begin{equation}
\text{ }\sigma _{xy}^{zz}=\frac{e}{{8}\pi },
\end{equation}%
when the Fermi level lays above the triply-degenerate point, while charge
and rank-1 spin Hall conductivities $\sigma _{xy}^{c}=\sigma _{xy}^{z}=0$.
We note that $\sigma _{xy}^{zz}=\frac{e}{{8}\pi }$ is a universal constant
that is independent of the material parameters such as SOC strength and
effective mass. Therefore the model Hamiltonian Eq.~(\ref{rank2H}) describes
a simple yet nontrivial system exhibiting universal higher-rank SHE.

To understand the physical mechanism of this rank-2 SHE, we notice that $%
\bm{\tau}_{T}\cdot (\bm{\hat{z}}\times \bm{p})$ yields{\ a} rank-1 SHE where 
$\left\vert +1\right\rangle $ and $|0\rangle $ spin components move in
opposite transverse directions (see Fig.~\ref{fig1}(b)), rendering a rank-1
spin-Hall conductivity $\sigma _{xy}^{T}=\frac{e}{8\pi }$. Accordingly, the
conjugate term $\bm{\tau}_{T}^{\ast }\cdot (\bm{\hat{z}}\times \bm{p})$
leads to an opposite spin flow, yielding a negative $\sigma _{xy}^{T^{\ast
}}=-\frac{e}{8\pi }$. While this can be verified by the Kubo formula, a more
insightful way is to look at the Bloch equation \cite{SinovaJ2004} 
\begin{equation}
\hbar \frac{d\bm{n}}{dt}=\bm{n}\times \bm{\Delta}+\eta _{d}\hbar \frac{d%
\bm{n}}{dt}\times \bm{n},
\end{equation}%
where $\bm{n}$ is the direction of the doublet, $\bm{\Delta}$ is the Zeeman
coupling (i.e. the SOC term) and $\eta _{d}$ is some small damping effects.
Considering the region where linear response theory applies and keeping only
leading-order terms, we have 
\begin{equation}
n_{z}=-\frac{e\hbar ^{2}E_{x}}{2\lambda }\frac{p_{y}}{p^{3}},
\end{equation}%
whose sign can be reversed by changing that of $p_{y}$. Notice that 
\begin{equation}
\bm{\tau}_{T}^{\ast }\cdot (\bm{\hat{z}}\times \bm{p})=\bm{\tau}_{T}\cdot
\left( \bm{\hat{z}}\times \left( \mathcal{I}\bm{p}\mathcal{I}^{-1}\right)
\right) ,
\end{equation}%
where $\mathcal{I}$ denotes spatial inversion under which $\bm{p}\rightarrow
-\bm{p}$. Therefore $n_{z}$ gains an opposite sign between Rashba SOC and
its conjugate term, leading to opposite rank-1 spin Hall conductivities.
Similar argument also applies to other subalgebras $\bm{\tau}_{U}$ and $%
\bm{\tau}_{V}$.

From above argument, $\bm{\tau}_{V}^{\ast }$ conjugate Rashba spin-orbit
coupling dictates that the $|0\rangle $/$\left\vert -1\right\rangle $
components flow along the same direction as the $|0\rangle $/$\left\vert
+1\right\rangle $ components under $\bm{\tau}_{T}$ (Fig.~\ref{fig1}(b)). 
{\ Therefore, Eq. (\ref{rank2H}) can be understood as coupled Rashba SOC and
its conjugate in different spin subspaces spanned, respectively, by $\left\{
\left\vert +1\right\rangle ,\left\vert 0\right\rangle \right\} _{T}$ and $%
\left\{ \left\vert 0\right\rangle ,\left\vert -1\right\rangle \right\} _{V}$
(Fig. \ref{fig1}(b)). The STMC term with $\bm{\tau}_{T}+\bm{\tau}_{V}^{\ast
} $ leads to the rank-2 spin-Hall conductivity $\sigma _{xy}^{zz}=\frac{e}{%
8\pi }$, while both rank-0 charge-Hall and rank-1 spin-Hall conductivities
vanish, as shown in Figs. \ref{fig2}(d) and \ref{fig2}(e). More intuitively,
the coupling between $\bm{\tau}_{T}$ and $\bm{\tau}_{V}^{\ast }$ terms
mediated by the shared $\left\vert 0\right\rangle $ guarantees the coherent
spin current of $\frac{1}{\sqrt{2}}$}$\left( {\left\vert +1\right\rangle
+\left\vert -1\right\rangle }\right) ${\ flowing in the opposite direction
to that of $\left\vert 0\right\rangle $. The coherent spin current and  the
rank-2 spin-Hall conductivity can also be derived under a gauge transform
that rotates the spin quantization axis from $z$ to $x$ direction (see
Supplementary materials \cite{SM}). }

Besides the rank-2 spin current $\mathbf{J}_{2}^{zz}$, we can similarly
define other higher-rank currents like $\mathbf{J}_{2}^{xy}$ or $\mathbf{J}%
_{2}^{yz}$, and the corresponding Hall conductivities usually vanish in the
above model. However, we notice the following constraint 
\begin{equation}
\sigma _{xy}^{xx}+\sigma _{xy}^{yy}+\sigma _{xy}^{zz}=0  \label{ConC}
\end{equation}%
holds as long as the charge current is zero. This constraint can be easily
proved since $\bm{F}^{2}$ must be a multiple of unit matrix (i.e., the total
spin is conserved). Applying the Kubo formula, we find 
\begin{equation}
\sigma _{xy}^{xx}=0\text{ and }\sigma _{xy}^{yy}=-\frac{e}{{8}\pi },
\end{equation}%
which indeed satisfies the above constraint.

When {$\bm{\tau}_{T}+\bm{\tau}_{V}^{\ast }$} in Eq.~(\ref{rank2H}) changes
to $\bm{\tau}_{T}+\bm{\tau}_{V}$, 
{\ the STMC transforms into the SOC, \textit{i.e.}, $-\frac{\lambda }{\hbar }%
\bm{F}\cdot (\hat{\bm{z}}\times \bm{p})$} \cite{HouJ2018}, and there is only
rank-1 SHE based on above argument, where the current of spin component $%
|0\rangle $ is cancelled out. Apply the Kubo formula, we find $\sigma
_{xy}^{z}=\frac{e}{2\pi }$ 
{, which counts the spin currents from the counterflow of both $\ket{\pm1}/%
\ket{0}$ and $\ket{+1}/\ket{-1}$.} Moreover, all rank-2 spin Hall
conductivities vanish so that Eq.~(\ref{ConC}) is trivially satisfied. 

{\color{blue}\emph{Effect of Zeeman fields}}. Generally, for a spin-1 system
discussed above, the Zeeman fields contain both spin-vector and spin-tensor
terms as{\ $\delta _{V}\hbar F_{z}+\delta _{T}\hbar F_{z}^{2}$}. Both fields
lift the triply-degenerate point at $\bm{p}=0$ and spoil the universality of
the spin Hall conductivity,{\ as shown in Fig. \ref{fig2}(b)-\ref{fig2}(e)}.
We numerically compute the{\ rank-1 and rank-2} spin-Hall conductivities,{\
which are displayed as a function of $\delta _{V}$ and $\delta _{T}$ in
Figs.~\ref{fig2}(d) and \ref{fig2}(e), respectively, for a given Fermi energy%
}.

Changing the sign of $\delta _{V}$ is equivalent to a $\mathbb{Z}_{2}$
rotation between spin components $|+1\rangle $ and $\left\vert
-1\right\rangle $, under which spin-tensor polarization ${\hbar }\langle
N_{zz}\rangle $ is unchanged while spin-vector polarization ${\hbar }\langle
F_{z}\rangle $ gains a minus sign, indicating symmetric and antisymmetric
responses from $\sigma _{xy}^{zz}$ and $\sigma _{xy}^{z}$, as shown in Fig.~%
\ref{fig2}(d). When $|\delta _{V}|$ increases, both $|\sigma _{xy}^{z}|$ and 
$|\sigma _{xy}^{zz}|$ increase to their maxima and then decrease. $\sigma
_{xy}^{zz}$ drops more rapidly as the top band shifts away from the Fermi
level. When $\delta _{V}\rightarrow \pm \infty $, both conductivities
approach 0 since the system becomes a flat-band insulator.

The tensor Zeeman field $\delta _{T}F_{z}^{2}$ has the same effect on $%
|+1\rangle $ and $\left\vert -1\right\rangle $, therefore $\sigma _{xy}^{zz}$
is asymmetric while $\sigma _{xy}^{z}$ remains zero [see Fig.~\ref{fig2}%
(e)]. There would also be small non-zero charge-current conductance, which
is not plotted in the panel. When $\delta _{T}>0$, it shifts both top and
bottom bands upward and they 
{\ remain degenerate at $\bm{p}=0$, as shown in Fig. \ref{fig2}(c). When $%
\delta _{T}>E_{F}$, the upper two bands are lifted above the Fermi energy,
while the three bands remain intersecting the Fermi level for $\delta _{T}<0$%
. Therefore, the rank-2 spin-Hall conductivity $\sigma _{xy}^{zz}$ decays
faster in the positive branch than that in the negative branch, as shown in
Fig. \ref{fig2}(e).} 

{\color{blue}\emph{Generalization to larger spin}}. A rank-2 SHE can also be
realized in a spin-$3/2$ system. Consider a 2D spin-3/2 Hamiltonian 
\begin{equation}
H_{F=\frac{3}{2}}=\frac{p^{2}}{2m^{\ast }}-\frac{\lambda }{\hbar }(\bm{\tau}%
_{1,2}+\bm{\tau}_{3,4}^{\ast })\cdot (\bm{\hat{z}}\times \bm{p}),
\label{H-3/2}
\end{equation}%
where $\bm{\tau}_{1,2}$ and $\bm{\tau}_{3,4}$ represent SU(2) subalgebra for 
$|m_{z}>0\rangle $ and $|m_{z}<0\rangle $ respectively \cite{SM}. The
Hamiltonian is block-diagonalized since the spin components $|m_{z}>0\rangle 
$ or $|m_{z}<0\rangle $ are coupled separately. The Hamiltonian describes
two decoupled rank-1 SHEs defined by $\tau _{1,2,z}$ and $\tau _{3,4,z}$ in
two subspaces. Specifically, spin components $|m_{z}\rangle $ and $%
\left\vert -m_{z}\right\rangle $ flow along the same direction, which is
opposite to that of $|m_{z}\pm 2\rangle $ ($-$ for $m_{z}>0$ and $+$ for $%
m_{z}<0$). Consequently, there is only non-vanishing rank-2 spin current
with Hall conductivity $\sigma _{xy}^{zz}=\frac{e}{2\pi }$.

Higher-rank SHE for arbitrary spin-$F$ can be defined similarly using the
algebra of SU($N=2F+1$) group (see supplementary material {\cite{SM})}. The
maximum rank of SHE is $N-1$, and spin current is defined using the
higher-rank spin tensor similar as $N_{zz}$. For a general SU($N$) group, we
can use the generalized Gell-Mann matrices \cite{GGM}, from which the SU(2)
subalgebras could be defined. Hamiltonians for realizing different ranks of
SHEs may be constructed similarly using the SOC based on such SU(2)
subgroups.

\begin{figure}[t]
\centering
\includegraphics[width=0.48\textwidth]{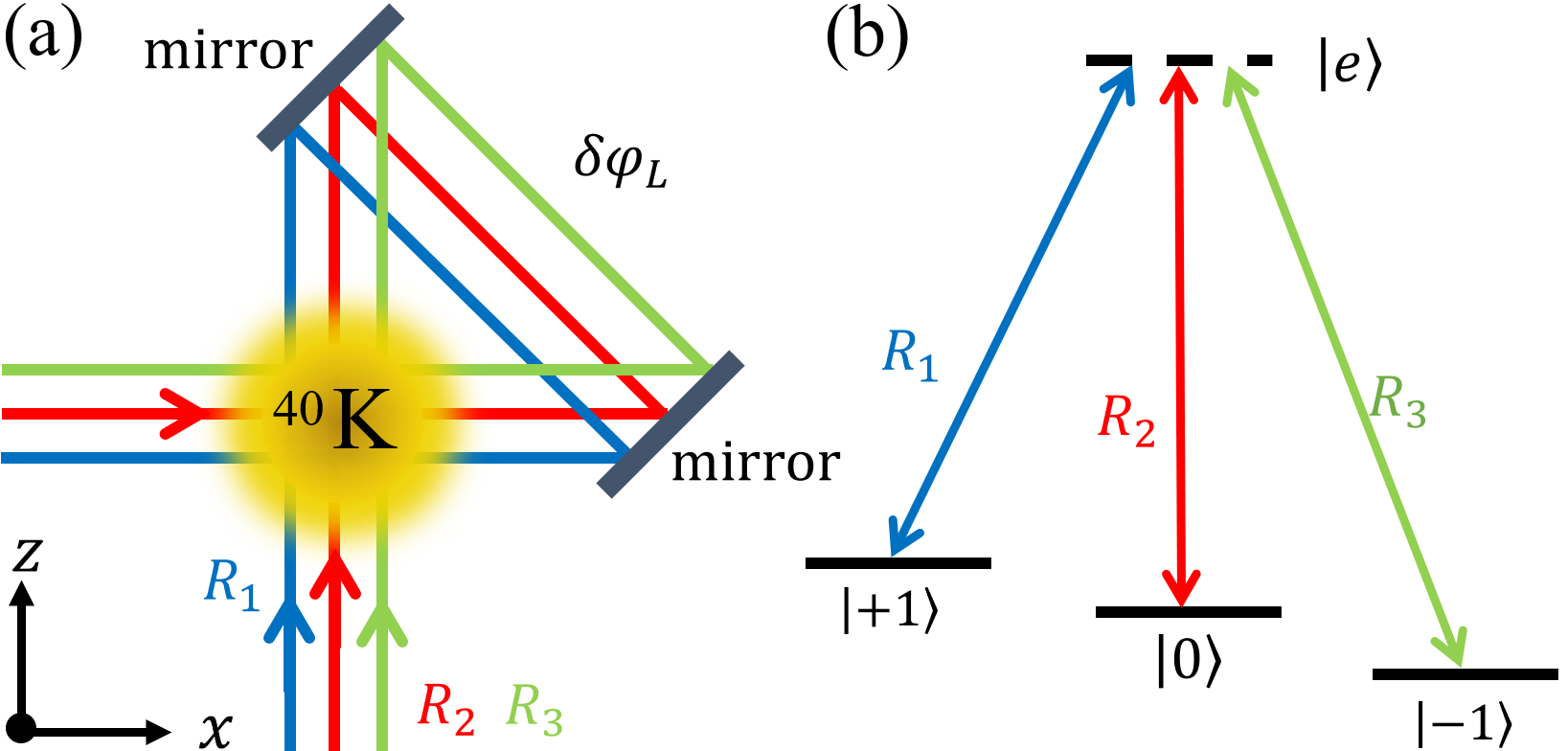}
\caption{Illustration of the experimental scheme for realizing STMC for the
rank-2 SHE in Eq.~(\protect\ref{rank2H}). (a) Laser setup consisting of
three Raman lasers $R_{1}$ (Blue, plane-wave), $R_{2}$ (red, standing wave)
and $R_{3}$ (green, plane wave). $\protect\varphi _{L}$ denotes the phase
accumulated by lasers in the triangle optical path $K$ formed by the atomic
gas and two mirrors. Then $\protect\delta \protect\varphi _{L}$ is the
differences between the accumulated phase. (b) The coupling between three
spin states under the Raman lasers with proper frequencies.}
\label{fig3}
\end{figure}

{\color{blue}\emph{Potential experimental realization and detection}}.
Recently, 2D SOC for spin-1/2 systems has been experimentally realized in
cold atoms \cite{WuZ2016,HuangLH2016,Meng2016}. Here we propose that similar
experimental setup \cite{WuZ2016} can also be used to realize the STMC in
the model Hamiltonian Eq.~(\ref{rank2H}) using a pseudospin-1 ultracold
atomic gas.

We use three hyperfine states of atoms to define (pseudo)spin states $%
\left\vert +1\right\rangle $, $\left\vert 0\right\rangle $, and $\left\vert
-1\right\rangle $. As demonstrated in Ref. \cite{WuZ2016}, either $\bm{\tau}%
_{T}\cdot (\bm{\hat{z}}\times \bm{p})$ or $\bm{\tau}_{V}^{\ast }\cdot (%
\bm{\hat{z}}\times \bm{p})$ can be realized\ in subspaces $\left\{
\left\vert +1\right\rangle ,\left\vert 0\right\rangle \right\} _{T}$ or $%
\left\{ \left\vert 0\right\rangle ,\left\vert -1\right\rangle \right\} _{V}$
by applying the Raman coupling between two spin states using a standing-wave
and a plane-wave Raman lasers. For the experiment from Ref. \cite{WuZ2016},
different forms of the SOC are tuned by adjusting a phase term $\delta
\varphi _{L}$, which is the accumulated relative phase between two Raman
beams when both travel through a given optical path. $\bm{\tau}_{T} $ and $%
\bm{\tau}_{T}^{\ast }$ terms correspond to $\delta \varphi _{L}=\frac{\pi }{2%
}$ and $-\frac{\pi }{2}$, respectively.

In order to generate coherent superposition of two different spin-orbit
couplings in different subspaces, two plane-wave and one standing-wave Raman
lasers are used to couple $\left\{ \left\vert +1\right\rangle ,\left\vert
0\right\rangle \right\} _{T}$ and $\left\{ \left\vert 0\right\rangle
,\left\vert -1\right\rangle \right\} _{V}$, respectively, as illustrated in
Fig. \ref{fig3}. Because of different forms of SOCs $\bm{\tau}_{T}$ and $%
\bm{\tau}_{V}^{\ast }$ in different subspaces, we need $\delta \varphi _{L}=-%
\frac{\pi }{2}$ for $\left\{ \left\vert +1\right\rangle ,\left\vert
0\right\rangle \right\} _{T}$ and $\frac{\pi }{2}$ for $\left\{ \left\vert
0\right\rangle ,\left\vert -1\right\rangle \right\} _{V}$ using the same
optical path. However, the phase tuning is unrealistic here due to the small
frequency difference between $R_{1}$ and $R_{3}$. We circumvent this issue
by choosing the laser frequency configuration in Fig. \ref{fig3}(b). Here
the blue plane-wave $R_{1}$ and red standing wave $R_{2}$ induce the Raman
coupling $\sim \Omega _{R_{1}}\Omega _{R_{2}}^{\ast }$ between $\left\vert
+1\right\rangle $ and $\left\vert 0\right\rangle $, while the red standing
wave $R_{2}$ and green plane-wave $R_{3}$ induce the coupling $\sim \Omega
_{R_{2}}\Omega _{R_{3}}^{\ast }$ between $\left\vert 0\right\rangle $ and $%
\left\vert -1\right\rangle $. $\Omega _{R_{i}}$ is the Rabi frequency for
corresponding Raman laser $R_{i}$. When $\Omega _{R_{1}}=\Omega _{R_{3}}$,
the complex conjugate condition could be satisfied, which realizes the
desired STMC in Hamiltonian Eq. (\ref{rank2H}).

To trigger the rank-2 SHE, an effective electric field for driving cold
atoms into motion can be achieved by applying a bias potential. The rank-2
spin current generated by the counterflow of $\ket{0}$ and $\frac{1}{\sqrt{2}%
}\ket{+1}+\frac{1}{\sqrt{2}}\ket{-1}$ leads to the accumulation of the two
states at opposite lateral edges, as illustrated in Fig. \ref{fig1}(b). The
spin accumulation can be detected as the signature of the rank-2 SHE,
similar as that for rank-1 SHE \cite{Kato2004}. More details about the
experimental scheme can be found in the supplementary materials \cite{SM}.

{\color{blue}\emph{Discussions and Conclusions}}. There are other types of
SOC, besides the discussed Rashba-type $\bm{\tau}\cdot (\bm{\hat{z}}\times %
\bm{p})$, leading to higher-rank SHE as well. For instance, we can mix
Dresselhaus and Rashba types in a spin-1 Hamiltonian 
\begin{equation}
H_{\text{Mix}}=\frac{\bm{p}^{2}}{2m^{*}}-\frac{1}{\sqrt{2}}\frac{\lambda }{%
\hbar }\left( (\bm{\tau}_{T}\cdot (\bm{\hat{z}}\times \bm{p})+\bm{\tau}%
_{V}\cdot \bm{p}\right) ,
\end{equation}%
which yields a universal intrinsic rank-2 SHE with $\sigma _{xy}^{zz}=\frac{e%
}{ 8\pi }$. In general, the characterization and symmetry requirement of the
SOC for realizing higher-rank SHEs would be an interesting topic to study.

In conclusion, we introduce and characterize the concept of higher-rank SHE
in large spin systems and propose its experimental realization in cold
atomic systems. There are many physics remaining to be explored, such as the
general construction of rank-$n$ SHE in arbitrary spin systems, quantized
higher-rank SHE that will enrich the category of topological insulators \cite%
{Hasan2010}, the effects of many-body interaction or disorders, extrinsic
higher-rank SHE, experimental proposal in solid-state materials with exotic
effective pseudospin-1 fermions in 2D \cite{WangSS2018}, and the observation
of higher-rank SHE in the parameter space \cite{Tan18,Tan21}, etc. Our work
defines a new class of SHEs and opens the door for designing large-spin
devices with novel functionalities for spintronic applications.

\begin{acknowledgments}
\textbf{Acknowledgements}: We thank Fan Zhang for helpful discussion. This
work was supported by Air Force Office of Scientific Research
(FA9550-16-1-0387, FA9550-20-1-0220), National Science Foundation
(PHY-1806227), and Army Research Office (W911NF-17-1-0128).
\end{acknowledgments}


\widetext
\newpage
\clearpage
\onecolumngrid
\begin{center}
\textbf{\large Supplemental Material for "Intrinsic high-rank spin Hall effect"}
\end{center}

\setcounter{equation}{0}
\setcounter{figure}{0}
\setcounter{table}{0}
\setcounter{page}{1}
\makeatletter
\renewcommand{\theequation}{S\arabic{equation}}
\renewcommand{\thefigure}{S\arabic{figure}}
\renewcommand{\bibnumfmt}[1]{[S#1]}
\renewcommand{\citenumfont}[1]{S#1}

\begin{figure}[tbp]
\centering\includegraphics[width=0.5\textwidth]{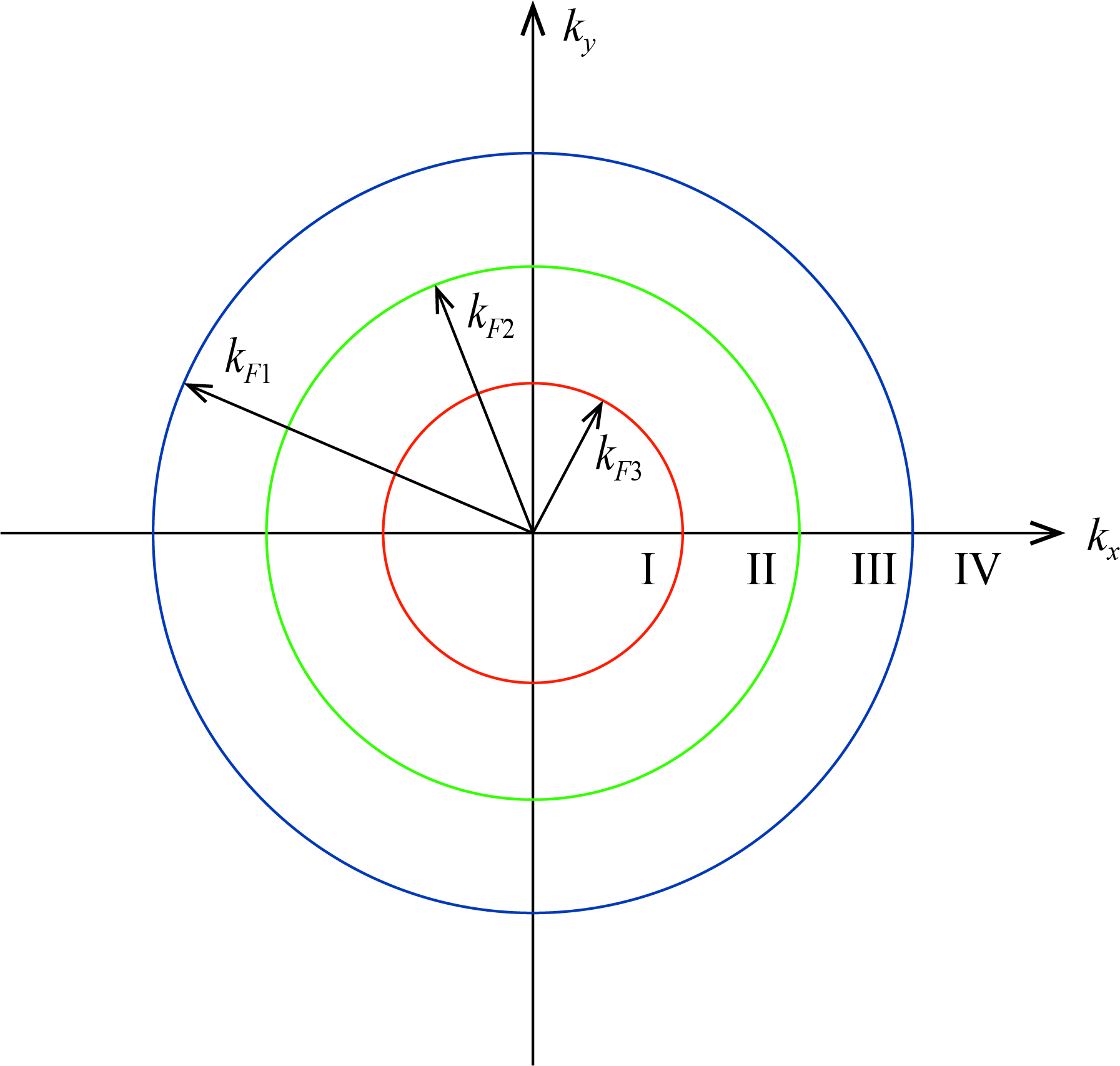}
\caption{Schematic\ of concentric Fermi surfaces. The momentum space is
divided into four different regions I-IV by the Fermi surfaces.}
\label{figs0}
\end{figure}

{\color{blue}\emph{S1: Derivation of the rank-2 spin-Hall conductivity}}.
Diagonalizing the Hamiltonian Eq. (4) of the main text yields three
eigenstates 
\begin{equation}
\ket{1\bm{k}}=\left( 
\begin{array}{c}
\frac{1}{2} \\ 
\frac{i}{\sqrt{2}}e^{i\theta _{\bm{k}}} \\ 
\frac{1}{2}%
\end{array}%
\right) ,\quad \ket{2\bm{k}}=\left( 
\begin{array}{c}
\frac{1}{\sqrt{2}} \\ 
0 \\ 
-\frac{1}{\sqrt{2}}%
\end{array}%
\right) ,\quad \ket{3\bm{k}}=\left( 
\begin{array}{c}
\frac{1}{2} \\ 
-\frac{i}{\sqrt{2}}e^{i\theta _{\bm{k}}} \\ 
\frac{1}{2}%
\end{array}%
\right) ,
\end{equation}%
where $\theta _{\bm{k}}$ is the polar angle of the wavevector $\bm{k}$, with
the corresponding eigenenergies 
\begin{equation}
E_{1\bm{k}}=\frac{\hbar ^{2}k^{2}}{2m}-\lambda k,\quad E_{2\bm{k}}=\frac{%
\hbar ^{2}k^{2}}{2m},\quad E_{3\bm{k}}=\frac{\hbar ^{2}k^{2}}{2m}+\lambda k.
\end{equation}%
Then we substitute the eigenstates and eigenenergies into the Kubo formula
Eq. (3) to calculate the rank-2 spin-Hall conductivity $\sigma _{xy}^{zz}$.
Here we consider zero temperature such that the Fermi-Dirac distribution is
a step function. The momentum space is divided into four different regions
I-IV that are separated by the three concentric Fermi surfaces, as shown in
Fig. \ref{figs0}. In region I (IV), all energy bands are occupied (empty)
with the Fermi-Dirac function $f_{1,2,3\bm{k}}=1$ ($f_{1,2,3\bm{k}}=0$).
Therefore regions I and IV have no contribution to the rank-2 spin-Hall
conductivity according to the Kubo formula Eq. (3). In region II, energy
bands $E_{1,2\bm{k}}$ are occupied while band $E_{3\bm{k}}$ is empty. In
region III, energy band $E_{1\bm{k}}$ is occupied while bands $E_{2,3\bm{k}}$
are empty. Thus the Kubo formula becomes 
\begin{equation}
\begin{split}
\sigma _{xy}^{zz}& =\frac{e\hbar }{4\pi ^{2}}\sum_{m=1,2}\int_{\mathrm{II}%
}d^{2}\bm{k}\frac{2\text{Im}\langle m\bm{k}|J_{2,x}^{zz}|3\bm{k}\rangle
\langle 3\bm{k}|\text{v}_{y}|m\bm{k}\rangle }{(E_{3\bm{k}}-E_{m\bm{k}})^{2}}+%
\frac{e\hbar }{4\pi ^{2}}\sum_{m=2,3}\int_{\mathrm{III}}d^{2}\bm{k}\frac{2%
\text{Im}\langle 1\bm{k}|J_{2,x}^{zz}|m\bm{k}\rangle \langle m\bm{k}|\text{v}%
_{y}|1\bm{k}\rangle }{(E_{m\bm{k}}-E_{1\bm{k}})^{2}} \\
& =\frac{e\hbar }{4\pi ^{2}}\int_{\mathrm{II+III}}d^{2}\bm{k}\frac{\hbar }{%
4M\lambda }\frac{k_{x}^{2}}{k^{3}}=\frac{e\hbar ^{2}}{16\pi ^{2}M\lambda }%
\int_{k_{F3}}^{k_{F1}}dk\int d\theta \cos ^{2}\theta \\
& =\frac{e\hbar ^{2}}{16\pi M\lambda }(k_{F1}-k_{F3})=\frac{e}{8\pi },
\end{split}
\label{sxy}
\end{equation}%
where $k_{F1,3}$ are the Fermi wavevectors, as shown in Fig. \ref{figs0},
and $k_{F1}-k_{F3}=2M\lambda /\hbar ^{2}$. In the first line of Eq. (\ref%
{sxy}), the factor 2 in front of the integrand comes from the exchange of
band indices in the Kuko formula. In the second line, the integration is
transformed into the polar coordinate. Our analytical derivation shows that,
as long as the Fermi energy is above the triply degenerate point, the rank-2
spin-Hall conductivity is quantized.

{\color{blue}\emph{S2: Spin dynamics analysis}}.{\ Because the
spin-tensor-momentum coupling (STMC) in Eq. (4) involves both spin tensor
and vector, we cannot treat it as an effective Zeeman interaction as that in
the Rashba spin orbit coupling. Consequently, the spin magnitude is not
conserved, as shown in Fig. 2(a), and the spin dynamics analysis cannot be
directly applied. In this section, we develop an effective theory for the
spin dynamics by performing} a gauge transformation that rotates the spin
quantization axis from $z$ to $x$ direction by the unitary transformation 
\begin{equation}
UF_{x}U^{-1}=F_{z},\quad UF_{y}U^{-1}=-F_{y},\quad UF_{z}U^{-1}=F_{x},
\end{equation}%
where the unitary matrix%
\begin{equation}
U=\left( 
\begin{array}{ccc}
\frac{1}{2} & \frac{1}{\sqrt{2}} & \frac{1}{2} \\ 
\frac{1}{\sqrt{2}} & 0 & -\frac{1}{\sqrt{2}} \\ 
\frac{1}{2} & -\frac{1}{\sqrt{2}} & \frac{1}{2}%
\end{array}%
\right) .
\end{equation}%
Thus the Hamiltonian Eq. (4) of the main text becomes 
\begin{equation}
\tilde{H}_{F=1}=UH_{F=1}U^{-1}=\frac{\bm{p}^{2}}{2m}+\frac{\lambda }{\hbar }%
(p_{y}\lambda _{+}+p_{x}\lambda _{5}),  \label{UHU}
\end{equation}%
in which the spin state $\ket{\tilde{0}}$ is decoupled with $%
\ket{\pm\tilde{1}}$ and has the dispersion $\bm{p}^{2}/2m$. 
Here $\ket{-\tilde{1}}=\frac{1}{2}\left( \left\vert +1\right\rangle
+\left\vert -1\right\rangle \right) -\frac{1}{\sqrt{2}}\left\vert
0\right\rangle $, $\ket{\tilde{0}}=\frac{1}{2}\left( \left\vert
+1\right\rangle -\left\vert -1\right\rangle \right) $, $\ket{\tilde{1}}=%
\frac{1}{2}\left( \left\vert +1\right\rangle +\left\vert -1\right\rangle
\right) +\frac{1}{\sqrt{2}}\left\vert 0\right\rangle $ are the spin states
quantized along the $x$ direction.

\begin{figure}[tbp]
\centering\includegraphics[width=0.6\textwidth]{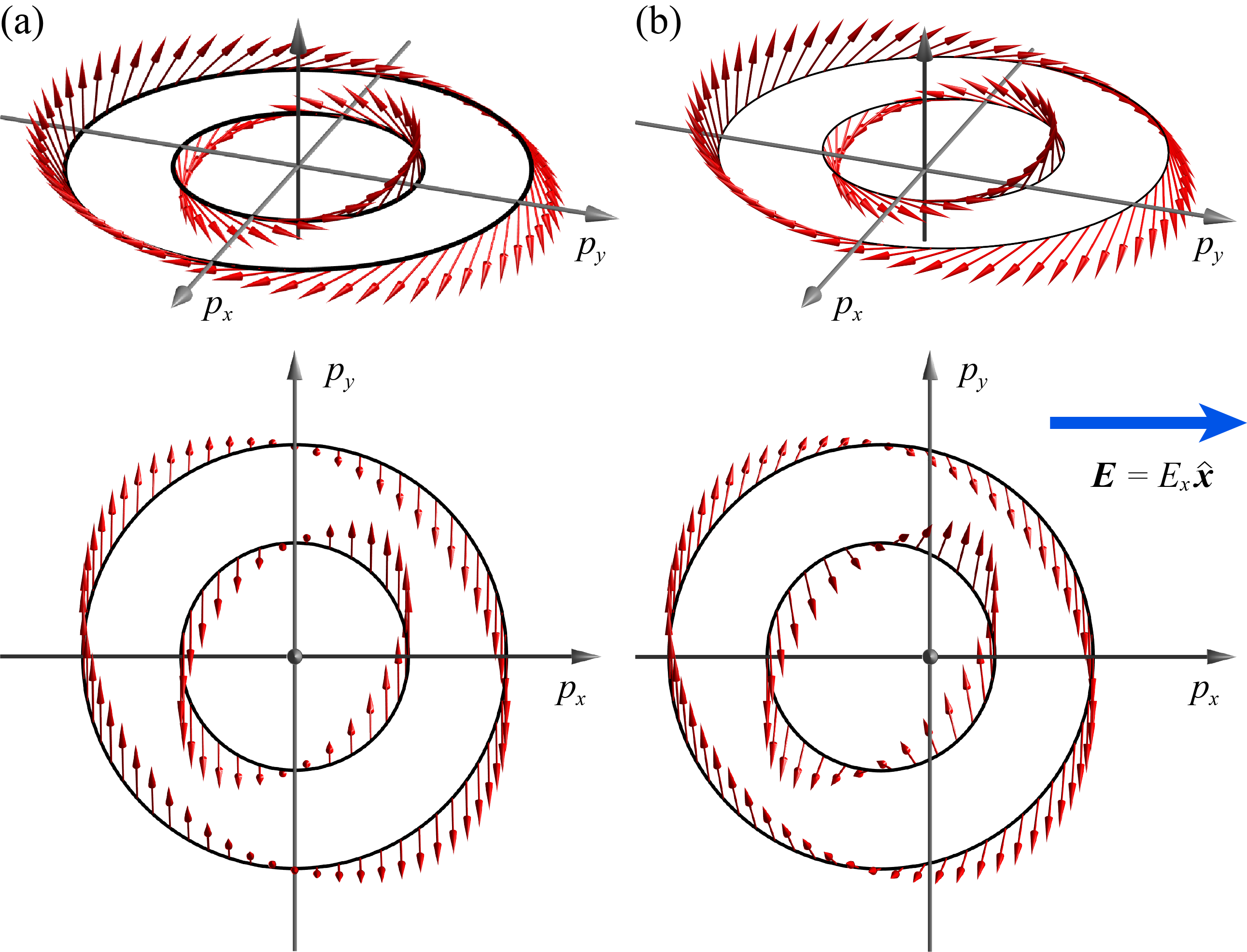}
\caption{(a) and (b) Spin texture on the Fermi surfaces with and without an
electric field in the $x$ direction. The upper panels show the bird view of
the spin texture, while the lower panels are the top view. The electric
field shifts the Fermi surfaces to the negative $p_{x}$ direction by $\sim
eE_{x}\protect\tau /\hbar $ where $\protect\tau $ is the mean free time. The
electric field tilts the spins with positive (negative) $p_{y}$ on the Fermi
surfaces to the positive (negative) $x$ direction.}
\label{figs1}
\end{figure}

The rank-2 spin tensor becomes 
\begin{equation}
\tilde{N}_{zz}=UN_{zz}U^{-1}=\left( 
\begin{array}{ccc}
\frac{1}{2} & 0 & \frac{1}{2} \\ 
0 & 1 & 0 \\ 
\frac{1}{2} & 0 & \frac{1}{2}%
\end{array}%
\right) -\frac{2}{3}I
\end{equation}%
under the transformation. Because $\bra{\tilde{0}}\mathbf{v}%
\ket{\pm\tilde{1}}=0$ where the velocity operator $\mathbf{v}=\partial _{%
\bm{p}}\tilde{H}$, the spin state $\ket{\tilde{0}}$ has no contribution to
the rank-2 spin-Hall conductivity, which can by directly inferred from the
Kubo formula Eq. (3) of the main text. Therefore, we can focus only on the
subspace spanned by $\{\ket{-\tilde{1}},\ket{+\tilde{1}}\}$. Now we project
the Hamiltonian and rank-2 spin tensor onto the subspace 
\begin{equation}
P\tilde{H}_{F=1}P=\frac{\bm{p}^{2}}{2m}+\frac{\lambda }{\hbar }(p_{y}\sigma
_{z}+p_{x}\sigma _{y}),  \label{PHP}
\end{equation}%
\begin{equation}
P\tilde{N}_{zz}P=\frac{1}{2}\sigma _{x}-\frac{1}{6}\sigma _{0},  \label{PNP}
\end{equation}%
where $P=\ket{+\tilde{1}}\bra{+\tilde{1}}+\ket{-\tilde{1}}\bra{-\tilde{1}}$
is the projection operator. According to Eq. (\ref{PHP}), the spin dynamics
are governed by the Bloch equation Eq. (6) of the main text with the $\bm{p}$%
-dependent Zeeman field $\bm{\Delta}=\frac{2\lambda }{\hbar }%
(0,-p_{x},-p_{y})$. Under an electric field along the $x$ direction, the $y$
component of the Zeeman field changes as $-\dot{p_{x}}=eE_{x}$. Solving the
Bloch equation yields the $x$ component of the spin direction that depends
on $\bm{p}$ as 
\begin{equation}
\tilde{n}_{x}=\frac{e\hbar ^{2}E_{x}}{2\lambda }\frac{p_{y}}{p^{3}}.
\label{nxp}
\end{equation}%
Because $\tilde{n}_{x}$ is an odd function of $p_{y}$, there is a transverse
rank-2 spin current $J_{2,y}^{zz}=\frac{1}{2}\{\hbar P\tilde{N}%
_{zz}P,v_{y}\}_{+}$. The electric field tilts the spin texture on the Fermi
surface according to Eq. (\ref{nxp}), as shown in Fig. \ref{figs1}. 
{Note that the spin texture here is different from that in Fig. 2(a) because
it is obtained by the gauge transformation and projecting out the $%
\ket{\tilde{0}}$ state.} The rank-2 SHE leads to the counterflow of currents
of $\frac{1}{\sqrt{2}}\ket{+\tilde{1}}\pm \frac{1}{\sqrt{2}}\ket{-\tilde{1}}$
which transform back into 
\begin{equation}
\begin{split}
& U^{-1}\left( \frac{1}{\sqrt{2}}\ket{+\tilde{1}}+\frac{1}{\sqrt{2}}%
\ket{-\tilde{1}}\right) =\frac{1}{\sqrt{2}}\ket{+{1}}+\frac{1}{\sqrt{2}}%
\ket{-{1}}, \\
& U^{-1}\left( \frac{1}{\sqrt{2}}\ket{+\tilde{1}}-\frac{1}{\sqrt{2}}%
\ket{-\tilde{1}}\right) =\ket{0},
\end{split}%
\end{equation}%
whose spin quantization axis is along the $z$ direction. Because both two
states have zero expectation value of $F_{z}$, the rank-1 spin-Hall
conductivity vanishes.

{\color{blue}\emph{S3: Higher-rank SHE for arbitrary spin-F}}. Mathematically,
it is well-known that a spin-$F$ Hamiltonian can be expanded using the
generators of the SU($N=2F+1$) group. Those generators are traceless and
symmetric, and can be constructed as rank-$n$ ($n\leq r_{s}$) spin tensors
from spin vectors $F_{i}$, $i=x,y,z$. Here $r_{s}=N-1$ is the rank of SU($N$%
) group. For instance, there are up to rank-2 spin tensors defined as $%
N_{ij}=\{F_{i},F_{j}\}_{+}/2-\delta _{ij}\bm{F}^{2}/3$ in a spin-1 system 
\cite{KawaguchiY2012}. We use a Cartan subalgebra $\{F_{z},N_{zz},\cdots
,N_{zz\cdots z}\}$ of SU($N$) to define a rank-$n$ spin polarization $P_{n}={%
\hbar }N_{zz\cdots z}$ (with $n$ subscripts) and spin current density $%
\mathbf{J}_{n}^{zz\cdots z}=\frac{1}{2}\left\{ P_{n},\mathbf{v}\right\} _{+}$%
operators. We can define rank-$n$ SHE as that with only non-zero rank-$n$
spin current (spin currents with different ranks, including charge current,
all vanish). Clearly, the first example is the rank-2 SHE, which requires at
least a spin-1 system.

For a general SU($N$) group, we can use the generalized Gell-Mann matrices 
\cite{GGM}, from which the SU(2) subalgebras could be defined. Consider a SU(%
$N$) Lie group, whose defining representation can be expressed as the
generalized Gell-Mann Matrix 
\begin{eqnarray}
\lambda _{j,k}^{S} &=&I_{j,k}+I_{k,j},\lambda _{j,k}^{A}=-i(I_{j,k}-I_{k,j}),
\\
\lambda _{l}^{D} &=&\sqrt{\frac{2}{l(l+1)}}\left(
\sum_{m=1}^{l}I_{m,m}-lI_{l+1,l+1}\right) ,  \notag
\end{eqnarray}%
where $1\leq j<k\leq N$, $1\leq l\leq N-1$ and $I_{j,k}$ denotes the matrix
with a 1 in the $(j,k)$-th entry and 0 elsewhere. We could similarly write
down all different SU(2) subalgebras as 
\begin{equation*}
\bm{\tau}_{j,k}=\{\lambda _{j,k}^{S},\lambda _{j,k}^{A},c_{j,k}^{l}\lambda
_{l}^{D}\},
\end{equation*}%
where $\bm{c}_{j,k}$ is some real vector so that the structure constant of
the SU(2) subalgebra is maintained.

Similar as the main text for the rank-2 SHE, Hamiltonians for realizing
different ranks of SHEs can be constructed using the SOC based on such SU(2)
subgroups. Since the Cartan subalgebra is complete up to a constant, this
configuration implies that there must be a non-vanishing rank-2 spin
current. Such a completeness also guarantees that a rank-$n$ SHE can always
be defined.

\begin{figure}[b]
\centering\includegraphics[width=0.8\textwidth]{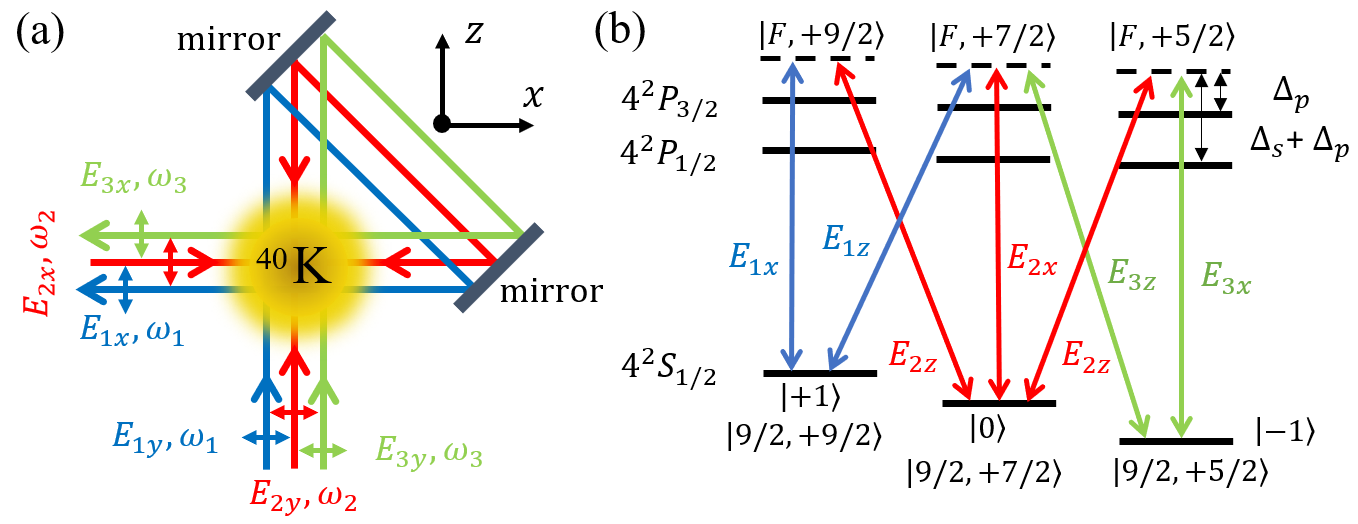}
\caption{(a) Experimental scheme for realizing spin-tensor-momentum coupling
of rank-2 SHE in three-component fermionic $^{40}$K atomic gases. The
desired 2D SO coupling is realized through a standing wave $\bm{E}_{2x(z)}$
and two plane-wave $\bm{E}_{1(3)x(z)}$ laser fields. The arrows indicate the
incident directions of the laser beams and each beam is reflected by two
mirrors. (b) Level diagram and optical coupling in the hyperfine structure $%
|F,m\rangle $ of $^{40}$K atoms. $F$ is the quantum number of hyperfine
states and $\Delta _{s}$ denotes fine-structure splitting.}
\label{figExp}
\end{figure}

{\color{blue}\emph{S4: Experimental realization of STMC}}. In this section, we discuss
the experimental scheme for realizing required spin-tensor-momentum coupling
for rank-2 SHE in a pseudospin-1 fermionic cold gas. Both laser
configuration and level diagram are illustrated in Fig.~\ref{figExp} with
more details. As 2D spin-orbit coupling has been realized in both bosonic
and fermionic cold-atom platforms \cite{WuZ2016,HuangLH2016,LiuXJ2014}, the
proposed experimental scheme can be readily implemented with the
state-of-art experimental technologies.

In Fig.~\ref{figExp}(a), two beams (red) are incident from both $x$ and $z$
directions and reflected by two mirrors to form standing waves $\bm{E}_{2x}=%
\hat{z}E_{2x}e^{i(\varphi_{2x}+\varphi _{2z}+\varphi _{L})/2}\cos
(k_{0}x+\alpha )$ and $\bm{E}_{2z}=\hat{x}E_{2z}e^{i(\varphi _{2x}+\varphi
_{2z}+\varphi _{L})/2}\cos (k_{0}z+\beta )$, where $E_{2x(z)}$ is field
strength, $\varphi _{2x(z)}$ is the initial phase, $\varphi _{L}=k_{0}K$ is
the phase picked up from optical path $K$ and $\alpha (\beta )=(\varphi
_{2x(z)}-\varphi _{2z(x)}-\varphi _{L})/2$. Another two laser beams (blue
and green) are incident in $z$ direction as plane waves $\bm{E}_{1(3)z}=\hat{%
x}E_{1(3)z}e^{i(k_{0}z+\varphi _{1(3)})}$ and $\bm{E}_{1(3)x}=\hat{z}%
E_{1(3)x}e^{i(-k_{0}x+\varphi _{1(3)}+\varphi _{L}-\delta \varphi _{L1(3)})}$
with the initial phases $\varphi _{1(3)}$ and relative phases $\delta
\varphi _{L1(3)}=(\omega _{2}-\omega _{1(3)})K/c$.

A similar scheme has been studied in our previous work \cite{HouJ2018},
where a low-energy Hamiltonian $k_{z}F_{x}+k_{x}F_{y}$ has been realized.
Such a model describes a rank-1 SHE as we have discussed in the main text.
The standing-wave laser will induce a spin-independent lattice potential in
this case 
\begin{equation}
V(\bm{r})=V_{0x}\cos ^{2}(k_{0}x+\alpha )+V_{0z}\cos ^{2}(k_{0}z+\beta ),
\end{equation}%
where $V_{0x(z)}$ are constants. The Raman coupling between $\left\vert
+1\right\rangle $ and $\left\vert 0\right\rangle $ can be written as 
\begin{equation*}
M_{1z,2x}=\sum_{F}\frac{\Omega _{1z,F,9/2}^{\ast }\Omega _{2x,F,7/2}}{\Delta
_{p}},M_{1x,2z}=\sum_{F}\frac{\Omega _{1x,F,9/2}^{\ast }\Omega _{2z,F,7/2}}{%
\Delta _{p}},
\end{equation*}%
where the effective Rabi frequency is 
\begin{eqnarray}
\Omega _{ix,F,m_{\sigma }} &=&e\langle \frac{9}{2},m_{\sigma }|z|F,m_{\sigma
}\rangle \hat{z}\cdot \bm{E}_{ix},i=1,2 \\
\Omega _{1z,F,m_{\sigma }} &=&e\langle \frac{9}{2},m_{\sigma }|x|F,m_{\sigma
}+1\rangle \hat{x}\cdot \bm{E}_{1z},~\Omega _{2z,F,m_{\sigma }}=\langle 
\frac{9}{2},m_{\sigma }|x|F,m_{\sigma }-1\rangle \hat{x}\cdot \bm{E}_{2z} 
\notag
\end{eqnarray}%
and we have neglected the transitions to $D1$ line due to larger
fine-structure splitting $\Delta _{s}\approx 2\pi \times 1.7\text{THz}\gg
\Delta _{p}$. After expanding the effective Rabi frequency, we obtain 
\begin{eqnarray}
M_{1z,2x} &=&M_{0x}\cos (k_{0}x+\alpha )e^{-i(k_{0}z+\beta )}e^{i(\varphi
_{2z}-\varphi _{1})}, \\
M_{1x,2z} &=&M_{0y}\cos (k_{0}z+\beta )e^{i(k_{0}x+\alpha )}e^{i(\varphi
_{2z}-\varphi _{1}+\delta \varphi _{L1})},
\end{eqnarray}%
and $M_{0x(y)}$ are coupling constants that can be tuned through individual
laser intensity. Similarly, the Raman coupling between $\left\vert
0\right\rangle $ and $\left\vert -1\right\rangle $ can be written as 
\begin{eqnarray}
M_{2z,3x} &=&\sum_{F}\frac{\Omega _{2x,F,7/2}^{\ast }\Omega _{3z,F,5/2}}{%
\Delta _{p}}=M_{0x}^{\prime }\cos (k_{0}x+\alpha )e^{i(k_{0}z+\beta
)}e^{i(-\varphi _{2z}+\varphi _{3})} \\
M_{2x,3z} &=&\sum_{F}\frac{\Omega _{2z,F,7/2}^{\ast }\Omega _{3x,F,5/2}}{%
\Delta _{p}}=M_{0y}^{\prime }\cos (k_{0}z+\beta )e^{-i(k_{0}x+\alpha
)}e^{i(-\varphi _{2z}+\varphi _{3}-\delta \varphi _{L3})}.
\end{eqnarray}%
We note that terms proportional to $\cos (k_{0}x+\alpha )\cos (k_{0}z+\beta
) $ are antisymmetric to each lattice site in both $x$ and $y$ directions
and thus can be neglected for low-band physics. The Raman coupling terms are
further simplified as 
\begin{eqnarray}
\mathcal{M}_{+1,0} &=&(M_{x}-M_{y}\cos \delta \varphi _{L1})-iM_{y}\sin
\delta \varphi _{L1}, \\
\mathcal{M}_{0,-1} &=&(M_{x}^{\prime }-M_{y}^{\prime }\cos \delta \varphi
_{L3})+iM_{y}^{\prime }\sin \delta \varphi _{L3},
\end{eqnarray}%
where $M_{x}=M_{0x}\cos (k_{0}x+\alpha )\sin (k_{0}z+\beta )$, $%
M_{y}=M_{0y}\cos (k_{0}z+\beta )\sin (k_{0}x+\alpha )$ and $M_{x(y)}^{\prime
}$ are defined similarly. In the above equations, we also choose the initial
phase of the lasers so that $e^{i(\varphi _{2z}-\varphi _{1})}=i$ and $%
e^{i(-\varphi _{2z}+\varphi _{3})}=-i$. Since $|\omega _{1}-\omega
_{3}|/\omega _{2}\ll 1$, we have $\delta \varphi _{L1}\approx \delta \varphi
_{L3}=\delta \varphi _{L}$. Assuming the coupling constants are tuned to be
equivalent $M_{x}^{\prime }=M_{x}$ and $M_{y}^{\prime }=M_{y}$, the total
effective Hamiltonian becomes 
\begin{equation}
H=\frac{\bm{p}^{2}}{2m}+V(\bm{r})+\mathcal{M}_{x}(\lambda _{1}+\lambda _{6})+%
\mathcal{M}_{y}(\lambda _{2}-\lambda _{7})+\frac{\delta _{T}}{2}F_{z}^{2}+%
\frac{\delta _{V}}{2}F_{z},
\end{equation}%
where $\mathcal{M}_{x}=M_{x}-M_{y}\cos \delta \varphi _{L}$, $\mathcal{M}%
_{y}=M_{y}\sin \delta \varphi _{L}$ and the Zeeman terms are incorporated
into the detunings in the ground-state manifold. When $\delta \varphi
_{L}=\pi /2$, the spin-orbit coupling becomes ${M}_{x}(\lambda _{1}+\lambda
_{6})+{M}_{y}(\lambda _{2}-\lambda _{7})$.

As we consider the lowest $s$-orbital $\phi _{s,\sigma }$ ($\sigma =+1,0,-1$%
) and nearest-neighbor hopping, the spin-orbit coupling part of the
tight-binding Hamiltonian can be written as 
\begin{equation*}
H_{\text{SOC}}=\sum_{\langle \bm{i},\bm{j}\rangle }\Big(t_{\text{so},+}^{%
\bm{ij}}\hat{c}_{\bm{i},+1}^{\dagger }\hat{c}_{\bm{j},0}+h.c.+t_{\text{so}%
,-}^{\bm{ij}}\hat{c}_{\bm{i},0}^{\dagger }\hat{c}_{\bm{j},-1}+h.c.\Big),
\end{equation*}%
where hopping strengths can be expressed as overlap integrals 
\begin{eqnarray}
t_{\text{so},+}^{\bm{ij}} &=&\int d^{2}\bm{r}\phi _{s,+1}^{\bm{i}}(\bm{r})%
\left[ M_{x}(x,y)(\lambda _{1}+\lambda _{6})+M_{y}(x,y)(\lambda _{2}-\lambda
_{7})\right] \phi _{s,0}^{\bm{j}}(\bm{r}), \\
t_{\text{so},-}^{\bm{ij}} &=&\int d^{2}\bm{r}\phi _{s,0}^{\bm{i}}(\bm{r})%
\left[ M_{x}(x,y)(\lambda _{1}+\lambda _{6})+M_{y}(x,y)(\lambda _{2}-\lambda
_{7})\right] \phi _{s,-1}^{\bm{j}}(\bm{r}).
\end{eqnarray}%
Finally, the spin-orbit coupling in low-energy Bloch Hamiltonian reads $%
\lambda _{\text{SO}}k_{x}(\lambda _{1}+\lambda _{6})+\lambda _{\text{SO}%
}k_{z}(\lambda _{2}-\lambda _{7})$, $\lambda _{\text{SO}}=2t_{\text{SO}}$,
which realizes the desired spin-tensor-momentum coupling for rank-2 SHE
discussed in main text.

{\color{blue}\emph{S5: Rank-2 spin accumulation.}} In the rank-2 SHE, the
counterflow of spin currents of $\ket{0}$ and $\frac{1}{2}\ket{+1}+\frac{1}{%
\sqrt{2}}\ket{-1}$ results in the rank-2 spin accumulation on the lateral
edges of the sample, as illustrated in Fig. 1 of the main text. Now we
consider cold atoms confined by a harmonic trap along the $y$ direction that
can be described by 
\begin{equation}
H=\tilde{H}_{F=1}+\frac{1}{2}m\omega ^{2}y^{2},
\end{equation}%
where $\tilde{H}_{F=1}$ encodes the spin-tensor-momentum coupling, as shown
in Eq. (\ref{UHU}). Here we assume the harmonic trap along the \textit{x}
direction is very weak and can be neglected in our calculation (or a box
potential along \textit{x }direction is considered). Because the harmonic
trap breaks the translational symmetry in the $y$ direction, we separate the
Hamiltonian into two parts $H=H_{0}+H_{1}$ where 
\begin{equation}
\begin{split}
& H_{0}=-\frac{\hbar ^{2}\partial _{y}^{2}}{2m}+\frac{1}{2}m\omega ^{2}y^{2}+%
\frac{\hbar ^{2}k_{x}^{2}}{2m}+\frac{\hbar ^{2}k_{x}k_{0}}{m}\lambda _{5}, \\
& H_{1}=-\frac{i\hbar ^{2}k_{0}\partial _{y}}{m}\lambda _{+},
\end{split}
\label{H0}
\end{equation}%
with $k_{0}=\lambda m/\hbar ^{2}$. $H_{0}$ is just the quantum oscillator up
to some constant terms. To simplify the notation, we set the oscillator
length $l_{\omega }=\sqrt{\frac{\hbar }{m\omega }}$ as length unit and $%
\frac{1}{2}\hbar \omega $ as energy unit. Then Eq. (\ref{H0}) becomes
dimensionless as 
\begin{equation}
\begin{split}
& H_{0}=-\partial _{\xi }^{2}+\xi ^{2}+l_{\omega }^{2}k_{x}^{2}+2l_{\omega
}^{2}k_{y}k_{0}\lambda _{5}, \\
& H_{1}=-2il_{\omega }k_{0}\partial _{\xi }\lambda _{+},
\end{split}%
\end{equation}%
where $\xi =y/l_{\omega }$.

\begin{figure}[tbp]
\centering\includegraphics[width=1\textwidth]{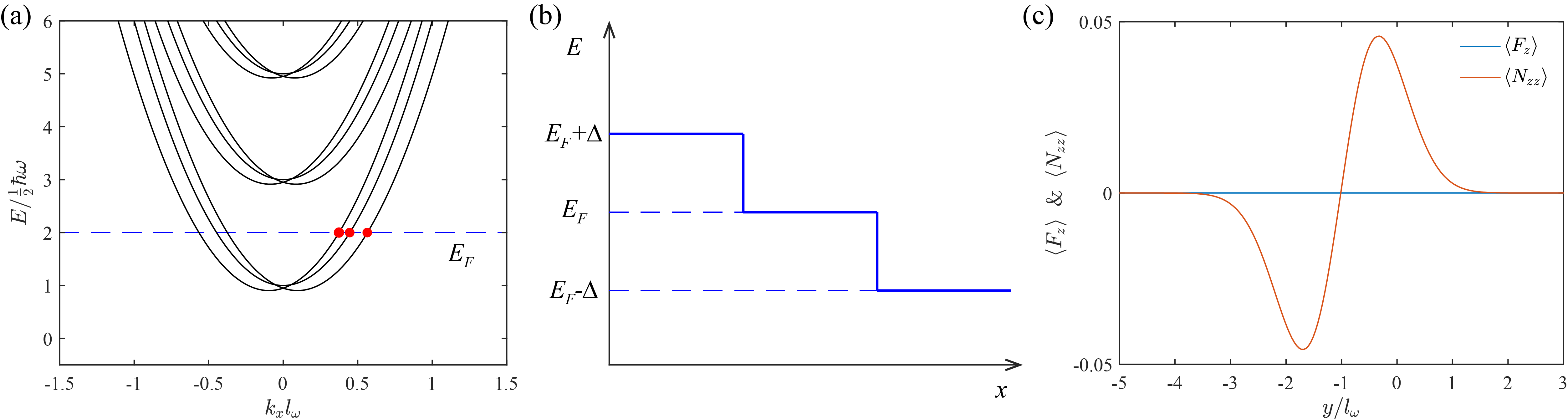}
\caption{(a) Energy spectrum of cold atoms with spin-tensor-momentum
coupling and confined by a harmonic trap in the $y$ direction. The blue
dashed line denotes the Fermi energy. (b) Schematic step potential applied
to drive cold atoms to move from left to right through the conducting
channels in the middle which are highlighted by red dots in (a). (c) The
distribution of rank-1 spin polarization $\langle F_{z}\rangle $ and rank-2
spin polarization $\langle N_{zz}\rangle $ in the middle region of (b) along
the $y$ direction. In the numerical simulation, we set $m=0.2$, $\protect%
\lambda =0.5$, $\protect\omega =1$, and $\hbar =1$. }
\label{figs3}
\end{figure}

The eigenvalue of $H_{0}$ is 
\begin{equation}
\varepsilon _{n,k_{x},s}=(2n+1)+l_{\omega
}^{2}(k_{x}+sk_{0})^{2}-s^{2}l_{\omega }^{2}k_{0}^{2}
\end{equation}%
and the corresponding wavefunction is 
\begin{equation}
\psi _{n,k_{x},s}(x,\xi )=e^{ik_{x}x}u_{n}(\xi )\chi _{s},\quad u_{n}(\xi )=%
\frac{1}{\sqrt{2^{n}n!\sqrt{\pi }}}e^{-\xi ^{2}/2}h_{n}(\xi ).
\end{equation}%
where $h_{n}(\xi )$ is the Hermite polynomial. $\chi _{s}$ is the spin part
of the wavefunction and the eigenstate of $\lambda _{5}$, whose eigenvalues
are $s=0,$ $\pm 1$. The wavefunction of the total Hamiltonian $H$ can be
expanded by $\psi _{n,k_{x},s}(x,\xi )$ as 
\begin{equation}
\Psi _{k_{x}}(x,\xi )=\sum_{n,s}c_{n,k_{x},s}\psi _{n,k_{x},s}(x,\xi ),
\end{equation}%
and then $H\Psi _{k_{x}}=E_{k_{x}}\Psi _{k_{x}}$ yields the coupled
equations 
\begin{equation}
(\varepsilon _{n,k_{x},+}-E_{k_{y}})c_{n,k_{x},+}+i\sqrt{2(n+1)}l_{\omega
}k_{0}c_{n+1,k_{x},-}-i\sqrt{2n}l_{\omega }k_{0}c_{n-1,k_{x},-}=0,
\end{equation}%
\begin{equation}
(\varepsilon _{n,k_{x},-}-E_{k_{y}})c_{n,k_{x},-}+i\sqrt{2(n+1)}l_{\omega
}k_{0}c_{n+1,k_{x},+}-i\sqrt{2n}l_{\omega }k_{0}c_{n-1,k_{x},+}=0,
\end{equation}%
\begin{equation}
(\varepsilon _{n,k_{x},0}-E_{k_{y}})c_{n,k_{x},0}=0.
\end{equation}%
Solving the three equations yields the energy spectrum of the system, as
shown in Fig. \ref{figs3}(a), and the corresponding wavefunctions.

To detect the spin accumulation, we need to drive the cold atoms to flow
along the $x$ direction, that can be achieved by applying, for instance, a
step potential (for the simplicity of the calculation), as shown in Fig. \ref%
{figs3}(b). In this configuration, the cold atoms move from the high
potential on the left to the low potential on the right through the
conducting channels in the middle region of Fig. \ref{figs3}(b). When the
bias energy $\Delta $ is much smaller than $\hbar \omega $, the conducting
channels are formed by states around the Fermi energy $E_{F}$, which are
highlighted by red dots in Fig. \ref{figs3}(a). Then we calculate the
distribution of rank-1 spin polarization $\langle F_{z}\rangle
=\sum_{i=1}^{3}\Psi _{i}^{\dagger }{\hbar }F_{z}\Psi _{i}$ and rank-2 spin
polarization $\langle N_{zz}\rangle =\sum_{i=1}^{3}\Psi _{i}^{\dagger }{%
\hbar }N_{zz}\Psi _{i}$ in the middle region, where $\Psi _{1,2,3}$ are the
wavefunctions of the three states highlighted by red dots in Fig. \ref{figs3}%
(b). Apparently, there is a rank-2 spin accumulation, as shown in Fig. \ref%
{figs3}(c), while the rank-1 spin accumulation vanishes.

\end{document}